\documentclass{article}
\usepackage{longtable}
\usepackage{booktabs}
\usepackage{amsmath}
\usepackage{mathtools}
\usepackage{algorithm, tabularx}
\usepackage{algpseudocode}
\usepackage{amssymb}
\usepackage{multirow}
\usepackage{hhline}
\usepackage{array}
\usepackage{makecell}
\usepackage{float}
\usepackage{chngcntr}
\usepackage{hhline}
\usepackage{caption}
\usepackage{subcaption}
\usepackage{bigints}
\usepackage{authblk}
\usepackage{hyperref}
\usepackage{multicol}
\usepackage{enumitem}

\setlist{noitemsep}

\newcommand{\eg}{{\it e.g.}}
\newcommand{\ie}{{\it i.e.}}

\newcommand{\ours}{\texttt{FaNC}}

\newcommand\norm[1]{\left\lVert#1\right\rVert}
\DeclareMathOperator*{\argmin}{arg\,min}

\newcommand{\beginsupplement}{
        \setcounter{table}{0}
        \renewcommand{\thetable}{S\arabic{table}}
        \setcounter{figure}{0}
        \renewcommand{\thefigure}{S\arabic{figure}}
        \setcounter{section}{0}
        \renewcommand{\thesection}{\Roman{section}}
        \setcounter{equation}{0}
        \renewcommand{\theequation}{S\arabic{equation}}
     }

\begin{document}

\title{Freudian and Newtonian Recurrent Cell for Sequential Recommendation}


\author[1,2]{Hoyeop Lee}
\author[1]{Jinbae Im}
\author[2,*]{Chang Ouk Kim}
\author[1]{Sehee Chung}
\affil[1]{Knowledge AI Lab., NCSOFT Co., South Korea}
\affil[2]{Department of Industrial Engineering, Yonsei University, South Korea}
\affil[*]{e-mail: kimco@yonsei.ac.kr}

\date{}

\maketitle

\begin{abstract}
A sequential recommender system aims to recommend attractive items to users based on behaviour patterns. 
The predominant sequential recommendation models are based on natural language processing models, such as the gated recurrent unit, that embed items in some defined space and grasp the user's long-term and short-term preferences based on the item embeddings.
However, these approaches lack fundamental insight into how such models are related to the user's inherent decision-making process.
To provide this insight, we propose a novel recurrent cell, namely \ours, from Freudian and Newtonian perspectives. 
\ours~divides the user's state into conscious and unconscious states, and the user's decision process is modelled by Freud's two principles: the pleasure principle and reality principle. To model the pleasure principle, \ie, free-floating user's instinct, we place the user's unconscious state and item embeddings in the same latent space and subject them to Newton's law of gravitation. Moreover, to recommend items to users, we model the reality principle, \ie, balancing the conscious and unconscious states, via a gating function.
Based on extensive experiments on various benchmark datasets, this paper provides insight into the characteristics of the proposed model. \ours~initiates a new direction of sequential recommendations at the convergence of psychoanalysis and recommender systems.
\end{abstract}


\section{Introduction}
\label{sec:intro}

With the explosion of information on the Internet, recommender systems~\cite{sarwar2001item,linden2003amazon,koren2009matrix} have become necessary for both users and service providers of web and mobile applications, \eg, web search, e-commerce, online music and video streaming, and social network services.
Most predominant recommender systems view estimating a sequence of user's preferences from the machine learning perspective and thereby use natural language processing (NLP) models, such as the recurrent neural network~\cite{cho2014learning}, attention mechanism~\cite{bahdanau2015neural}, and their variants~\cite{vaswani2017attention,devlin-etal-2019-bert}, to capture the user's preference dynamics and to predict subsequent items~\cite{hidasi2015session,twardowski2016modelling,he2017translation,kang2018self}. Such sequential models have shown remarkable performance in the various recommendation domains, such as movies~\cite{harper2015movielens}, music~\cite{van2013deep}, news articles~\cite{li2010contextual}, and e-commerce~\cite{mcauley2015image}, by encoding items in a latent space and modelling the dynamics of user's short-term and long-term preferences in the latent space based on the encoded items. 
However, such approaches lack fundamental insight into how the models are related to the user's inherent decision-making process.

\begin{figure*}[t]
    \centering
    \includegraphics[width=0.9\textwidth]{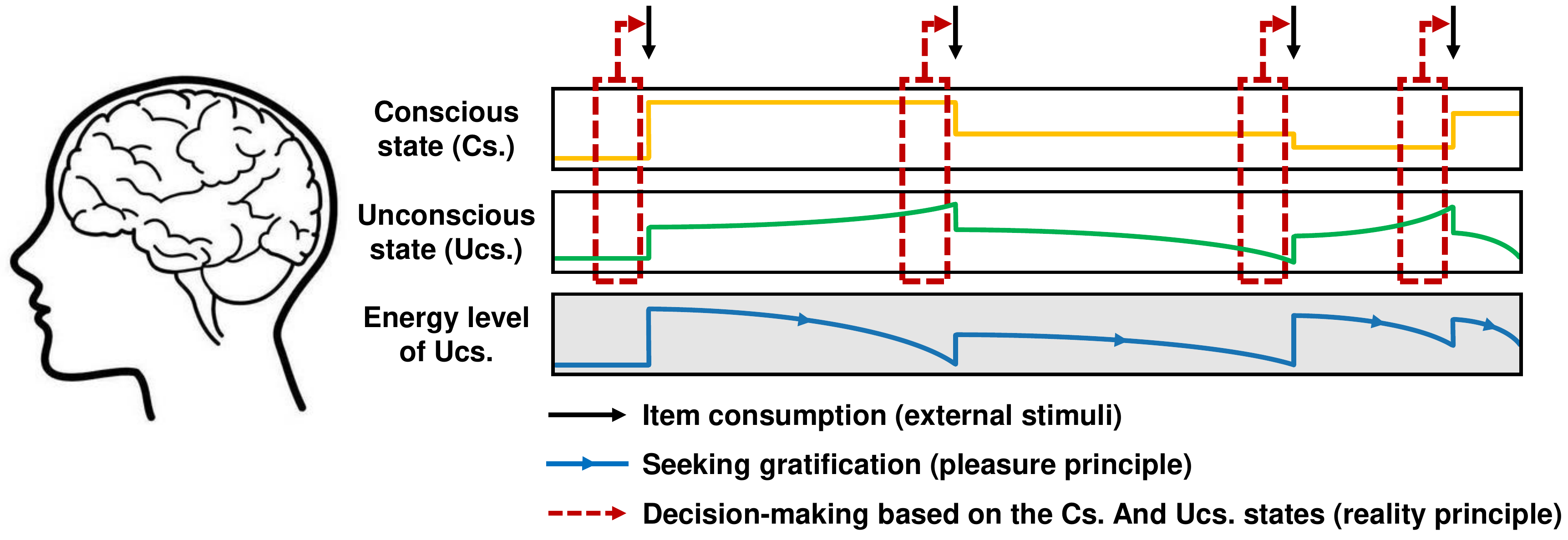}
    \caption[Illustration of the dynamics of the conscious and unconscious states and the consequent decision-making process.]{Illustration of dynamics of conscious and unconscious states and consequence decision-making process. 
    As Freud's theory states, the decision-making process is derived from the conscious and unconscious, so we divide a user's state into conscious and unconscious states. The conscious and unconscious states are modelled via the pleasure principle and reality principle. Before the two principles are applied, the purchased item (\ie, external stimuli) shifts the conscious and unconscious states.
    Afterwards, the pleasure principle enables the unconscious state to seek gratification by reducing its energy level.
    Specifically, we place the unconscious state and items in the same latent space so that there is a gravitational potential energy. 
    In other words, the unconscious state and items follow Newton's law of universal gravitation, which implies that an item attracts the user unconsciously.
    We regard the floating unconscious state according to gravity as the pleasure principle because the energy decreases as the unconscious state flows through gravity.
    Subsequently, the reality principle enables the final decision (\ie, select next item) by balancing the two shifted and free-floating states.
    }
    \label{fig:illustration}
\end{figure*}

This paper proposes a novel sequential recommendation model that provides insights from two perspectives: Freudian~\cite{freud1900interpretation} and Newtonian~\cite{newton1999principia}.
From the Freudian perspective, claiming that a human's decision-making process is derived from not only the \emph{conscious} but also the \emph{unconscious}~\cite{dijksterhuis2004think,newell2014unconscious},
we expect to enhance the recommendation model by reflecting the user's conscious and unconscious.
Freud, the founder of psychoanalysis, interpreted the conscious and unconscious decision-making process through two competing principles, the pleasure principle and reality principle~\cite{freud1958formulations}.
The pleasure principle encourages immediate gratification of unconscious instinct, but the instinct is interrupted by the reality principle, which represents mankind's conscious being \emph{rational}.
In this paper, the conscious state represents the user's thoughts directly affected by interactions, such as consumption, and we suppose that this state exists in a latent space.
Moreover, the unconscious state represents the \emph{free-floating} user's instinct underlying the conscious state, and this paper assumes that it lies in another latent space based on the dissociations between explicit (\ie, conscious) and implicit (\ie, unconscious) memory~\cite{poldrack2001interactive,rugg1998dissociation}.
To model the two competing principles for sequential recommendation, this paper regards buying a product and the subsequent experience with the product as an external stimulus, as shown in Fig.~\ref{fig:illustration}. 
Since the product and subsequent experience change our conscious perception of the product family, the external stimulus shifts the conscious state.
Afterwards, as per psychological findings that the conscious selectively affects the unconscious, \eg, by sleep~\cite{saletin2011role}, we let the conscious state discriminatively impact the unconscious state.
Thus, the external stimulus indirectly makes the unconscious state nervous.
In this paper, the pleasure principle lets the unconscious state be free-floating by reducing the energy level of the unconscious (\ie, psychic energy~\cite{freud1955beyond}), and the state becomes calm. 
Consequently, the reality principle allows decision making to be \emph{appropriate} for selecting a subsequent item by balancing the conscious and unconscious states.
In other words, the customer's next purchase is influenced by the shifted and free-floating states, which means `buying behaviour' and `conscious and unconscious states' affect each other.
Note that when explaining the buying decision process, we use the term `appropriate' rather than `rational' because not all users make rational consumption decisions, \eg, impulse buying~\cite{rook1987buying}.

The Freudian perspective's main challenge is modelling the fluidity of the unconscious state, and we approach this task from the Newtonian perspective.
To do so, we remind the reader of Newton’s law of universal gravitation: the attraction between two particles in Euclidean space is directly proportional to their masses and inversely proportional to the square of the distance between them~\cite{newton1999principia}. 
As a user is involuntarily attracted to an item, we presume that the user's unconscious state and items are the particles of Newton's law.
Specifically, we place the unconscious state and items in the same latent space, and we let the items be fixed in the latent space, similarly to stars on the celestial surface in traditional astrophysics. 
By contrast, our model enables the unconscious state to float through gravitational accelerations between the unconscious state and items during the item-consumption time intervals.
Thus, the unconscious state has gravitational potential energy, and the energy might increase (\ie, nervous) when external stimuli are given.
Since the potential energy decreases (\ie, calm) as the unconscious state flows according to gravity, this paper views the floating unconscious state via Newton's law as Freud's pleasure principle.

\subsection{Problem Statement}
A sequential recommender system aims to recommend attractive items to a user based on the user's behaviour in terms of recently consumed items. 
This statement can be mathematically formulated as follows.
Let $\mathcal{I}=\left\{i_{1}, \ldots, i_{N}\right\}$ be a set of items. 
In contrast to previous studies that denote a behaviour sequence as $\mathcal{S}=\left[ s_{1}, \ldots, s_{L}\right]$, we define the behaviour sequence with $L$ recent items as 
\begin{equation}
    \label{eq:session}
    \mathcal{S}=\left[ (s_{1}, t_1), \ldots, (s_{L}, t_L)\right],
\end{equation}
where $s_{l} \in \mathcal{I}\ \forall l \in \{1, \ldots, L \}$ represents the $l$-th interacted item index and $t_{l}$ is the interaction time with the item. 
Because the previous definition cannot consider time information, we modify the definition for our study, where the unconscious state is free-floating during the time interval.
The interaction behaviour can be purchasing, writing a review, or browsing. Notably, our definition of behaviour sequence includes the previous definition because it can be regarded as a special case of equation~(\ref{eq:session}), which corresponds to the case of $t_l=l\ \forall l\in\{1,\ldots,L\}$ (\ie, equal time intervals). 
Given the behaviour sequence $\mathcal{S}$ and the estimated next interaction time $t_{L+1}$, the sequential recommendation model $\mathcal{F}$
aims to recommend the next item $s_{L+1}$, and this process can be formalised as
\begin{equation}
    \label{eq:problem}
    s_{L+1} = \mathcal{F} \left( \mathcal{S}, t_{L+1} \right).
\end{equation}

Thus, the recommendation model expects the target user to interact with item $s_{L+1}$ at time $t_{L+1}$.
Our recommendation model allows us to provide a what-if analysis of consumption time. 
In other words, we can determine what item a user would have consumed if the user revisited in a week rather than a month. 
The answer is necessary to service providers because they can determine the time for promotion.
Although research on forecasting the next interaction time $t_{L+1}$~\cite{wang2013opportunity,bhagat2018buy} is worth conducting, we consider it as being out-of-scope of this paper.

\section{Model}
\label{sec:model}
\subsection{Overview}
\begin{figure*}[t]
    \centering
    \includegraphics[width=0.95\textwidth]{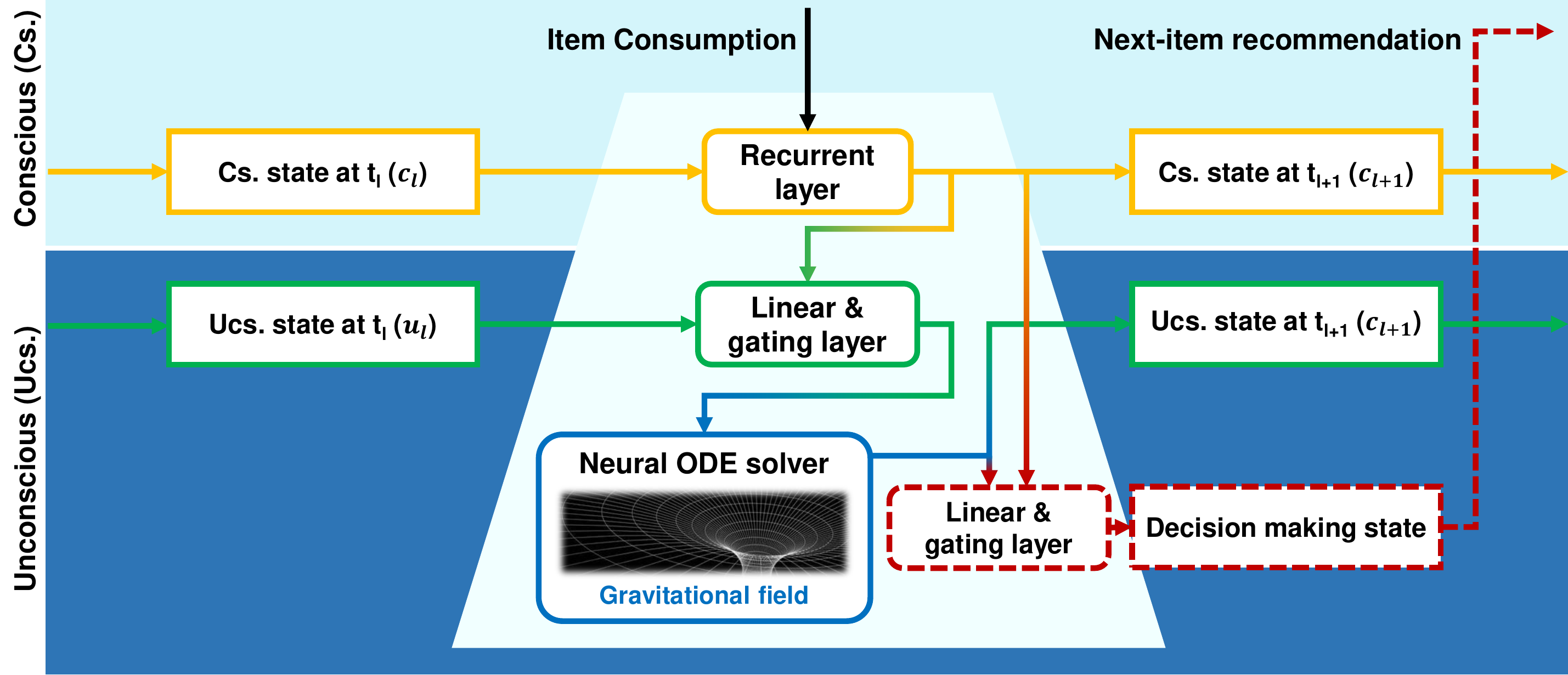}
    \caption[Architecture of the Freudian and Newtonian recurrent cell.]{Architecture of the Freudian and Newtonian recurrent cell, namely, \ours. To follow Freud's (iceberg) theory, we place the trapezoid cell in the shape of an iceberg between the blue sky and the blue ocean, which represent the areas of conscious and unconscious, respectively. The cell receives preceding conscious and unconscious states and a consumed item. Since the conscious state changes when an event occurs, \ie, item consumption, we design the process through the recurrent layer, which generates the following conscious state. \ours~modifies the previous unconscious state to the following unconscious state via the linear \& gating layer and neural ordinary differential equation (ODE) solver. The former implies that the conscious state selectively affects the unconscious state, and the latter is used to model the pleasure principle. Specifically, we employ Newton's law of universal gravitation and model it via the neural ODE solver to model the pleasure principle. Then, \ours~generates a decision-making state by balancing the following conscious and unconscious states via another linear \& gating layer that mimics the reality principle. Finally, our model provides a recommendation result based on the decision-making state.}
    \label{fig:proposedmodel}
\end{figure*}

This paper proposes a Freudian and Newtonian recurrent cell, namely, \ours~(pronounced ``fancy"), as illustrated in Fig.~\ref{fig:proposedmodel}. 
Generally, a sequential recommendation model $\mathcal{F}$ is composed of three layers: item embedding, sequence modelling, and recommendation layers~\cite{jang2020cities}. 
The item embedding layer encodes items into a low-dimensional real-valued latent space to measure the similarity between items.
Based on the embedded items, the sequence modelling layer captures the dynamics of the user's interests, which affect the decision-making process.
The recommendation layer represents the user's decision-making process and predicts which item to consume in accordance with the captured interest.
\ours~is a novel sequence modelling layer that models user's conscious and unconscious sequential behaviour by dividing the user's hidden state into conscious and unconscious states.
The sequential dynamics of the conscious state are modelled by the conventional recurrent layer.
To model the flow of the unconscious state via Newton's law of gratification (\ie, pleasure principle), we employ the neural ordinary differential equation (ODE) solver.
\ours~produces a decision-making state by balancing the conscious and unconscious states (\ie, reality principle), and it includes a user's interests.
Thus, the subsequent recommendation layer makes a recommendation based on the decision-making state.

\subsection{Item embedding layer}
As in NLP research~\cite{mikolov2013distributed,pennington2014glove}, recent studies in sequential recommendation convert an item index into a low-dimensional vector, called an item embedding vector.
Traditional recommendation models often use categorical metadata of items, such as genre and manufacturer, and convert the item index into a concatenation of multiple one-hot vectors in which only the item's metadata value is one and the remaining values are zero~\cite{debnath2008feature}.
These vectors are very sparse, and the similarity between items is difficult to measure.
Recent works~\cite{hidasi2015session,kang2018self,jang2020cities,sun2019bert4rec} employ an embedding layer to resolve these limitations, and this study also uses an item embedding layer.
The item embedding layer $f_\epsilon: \mathbb{N} \rightarrow \mathbb{R}^{d_u}$ projects the item index onto a $d_u$-dimensional real-valued dense vector as
\begin{equation}
    \label{eq:item_emb}
    e_l = f_\epsilon(s_l), \forall l
\end{equation}
where $e_l$ represents the item embedding vector for item $s_l$.
We assume that the item embedding vector is in the same space as the unconscious state so that we can apply Newton's law, which is why we use $u$ as a subscript of dimensionality.

\subsection{FaNC: A novel sequence modelling layer}
\ours~operates in the same way as a recurrent cell by receiving a consumed item and the previous hidden state and producing a new hidden state.
Since we divide the user's hidden state into conscious and unconscious states, our cell takes the consumed item embedding $e_l$ and previous conscious and unconscious states, denoted as $c_{l-1}$ and $u_{l-1}$, respectively, and outputs the new conscious and unconscious states, $c_{l}$ and $u_{l}$. Furthermore, \ours~generates a decision-making state $d_{l}$ by balancing the conscious and unconscious states. We introduce the detailed procedure of \ours~as follows.

\subsubsection{Modelling the dynamics of the conscious state}
\ours~models the dynamics of the user's conscious state $c_l \in \mathbb{R}^{d_c}$ with a recurrent layer that updates the conscious state only when the user consumes an item, where $d_c$ is the dimensionality of the conscious state. 
Specifically, the recurrent layer has the same architecture as the gated recurrent unit (GRU)~\cite{cho2014learning} because the GRU outperforms other recurrent cells for sequential recommendation~\cite{hidasi2015session}. 
The GRU is commonly represented as 
\begin{equation}
    c_l = \mathrm{GRU}(c_{l-1}, e_l).
\end{equation}
Additionally, we can expand the GRU function into the following equations~(\ref{eq:gru_basic})--(\ref{eq:gru_reset_gate}).
The GRU linearly interpolates the previous conscious state $c_{l-1}$ and the candidate conscious state $\hat{c}_{l}$ as
\begin{equation}
    \label{eq:gru_basic}
    c_l = (1-z_l) c_{l-1} + z_l \hat{c}_{l},
\end{equation}
where the update gate $z_l$ is given by
\begin{equation}
    \label{eq:gru_update_gate}
    z_l = \sigma(W_z e_l + U_z c_{l-1}),
\end{equation}
where $\sigma$ is an activation function, such as a logistic sigmoid function.
The candidate conscious state $\hat{c}_{l}$ is computed as follows:
\begin{equation}
    \label{eq:gru_candidate_function}
    \hat{c}_{l} = \mathrm{tanh} (W_c e_l + U_c(g_l \odot c_{l-1})),
\end{equation}
where $\odot$ represents the element-wise product and the reset gate $g_l$ is calculated as
\begin{equation}
    \label{eq:gru_reset_gate}
    g_l = \sigma(W_g e_l + U_g c_{l-1}).
\end{equation}

\subsubsection{Modelling the dynamics of the unconscious state}
We suppose that the free-floating dynamics of the unconscious state follow Freud's pleasure principle and model the dynamics via Newton's law of universal gravitation. 
Since we presume the item embedding vectors and unconscious state are in the same latent space, the unconscious state has gravitational potential energy.
This paper regards reducing the potential energy as imitating the pleasure principle.
To do so, we let the unconscious state $u_l$ be floating according to the gravitational acceleration between the unconscious state and the whole set of items $\mathcal{I}$ during the item-consumption time intervals $t_{l-1}$ to $t_l$. 
Based on physics research~\cite{elmer1982gravitational,wilkins1986gravitational}, the $d_u$-dimensional gravitational force $F$ applied to the unconscious state $u$ by an item $i$ at time $t$ is calculated as
\begin{equation}
    \label{eq:force}
    {F}(u,i,t)=\frac{Gm_u m_i}{\norm{{r}(u,i,t)}^{d_u}}{r}(u,i,t),
\end{equation}
where ${r}(u,i,t)$ is the displacement from $u$ to $i$, $G$ is the gravitational constant, $m_u$ and $m_i$ are the masses of the unconscious state and item $i$, and $\norm{\cdot}$ denotes the norm of a vector. The mass of item $m_i$ represents the attractiveness of item $i$.
Based on equation~(\ref{eq:force}), we calculate the net force, which is the sum of the forces of all items and determines the net acceleration, as
\begin{equation}
    \label{eq:netforce}
    \underbrace{\sum_{i\in\mathcal{I}}{{F}(u,i,t)}}_\text{\clap{net force~}}=m_u\underbrace{\sum_{i\in\mathcal{I}}{\frac{Gm_i }{\norm{{r}(u,i,t)}^{d_u}}{r}(u,i,t)}}_\text{\clap{~net acceleration}}=m_u{a}(u,t),
\end{equation}
where ${a}(u,t)$ is the net acceleration for unconscious state $u$ at time $t$. 
This paper does not address the mass of unconscious $m_u$ because it does not affect the position (\ie, unconscious state $u$). Since only the net acceleration influences the position, we can write the relation between position and net acceleration as the following second-order ODE:
\begin{equation}
    \label{eq:gravityode}
    \ddot{u}(t)={a}(u,t)
\end{equation}
Equation~(\ref{eq:gravityode}) can be reduced to an equivalent system of coupled first-order ODEs:
\begin{align}
    \label{eq:coupledode}
    \left\{
    \begin{aligned}
        \dot{u}(t)&=v(u,t) \\
        \dot{v}(u,t)&={a}(u,t)
    \end{aligned}\right., \qquad
    \begin{bmatrix}
         u_{l}\\
         v_{l}
    \end{bmatrix} &=
    \begin{bmatrix}
         u_{l-1}\\
         v_{l-1}
    \end{bmatrix}+
    \bigintsss_{t_{l-1}}^{t_{l}}\! \begin{bmatrix}
         v(u,t)\\
         {a}(u,t)
    \end{bmatrix} \mathrm{d} t, \\
    &=
    \begin{bmatrix}
         u_{l-1}\\
         v_{l-1}
    \end{bmatrix}+
    \bigintsss_{0}^{t_{l}-t_{l-1}}\! \begin{bmatrix}
         v(u,t'+t_{l})\\
         {a}(u,t'+t_{l})
    \end{bmatrix} \mathrm{d} t', \label{eq:substitue}
\end{align}
where $v_l$ is the unconscious state's velocity at time $t_l$. Furthermore, we call the combined vector of $u_l$ and $v_l$ the extended unconscious state, \ie, $h_l=[u_l,v_l]$. In addition, this paper sets $u_0$ and $v_0$ as zero vectors.
Additionally, ${a}(u,t)$ changes as the unconscious state $u$ varies in every moment because equation~(\ref{eq:netforce}) holds.
We substitute $t'$ for $t-t_l$ in equation~(\ref{eq:coupledode}) and obtain equation~(\ref{eq:substitue}), which implies that time does not matter when the position and velocity are the same in the gravitational field.
To solve the above ODEs, we have to determine $G$ and $m_i$ in equation~(\ref{eq:netforce}).
We set the gravitational constant $G$ to one in the recommender system for ease of calculation such that we need to estimate only the parameters $m_i$.

From the machine learning perspective, to estimate the parameters of the unconscious state under the ODE, this paper employs a neural ODE~\cite{chen2018neural}, which is the continuous deep learning model. The neural ODE considers the ODE solver as a black box function and conducts backpropagation using the adjoint method~\cite{pontryagin2018mathematical}. 
The black box function receives an initial extended unconscious state $h_{l-1}$, gravitational acceleration function ${a}(u,t)$, start time $t_{l-1}$, stop time $t_{l}$, and the parameters $m_i$ and $e_i$ for all $i$, and produces a floated extended unconscious state $h_{l}$, as follows:
\begin{equation}
    \label{eq:odesolver}
    h_{l} =\mathrm{ODESolver}\left(h_{l-1}, {a}(u,t), t_{l-1}, t_{l}, \{m_i, e_i | i \in \mathcal{I} \}\right).
\end{equation}

However, ODEs have a critical limitation that they cannot cross paths, which yields an important implication: modelling the unconscious state based on only ODEs indicates that most users have isolated preferences. Therefore, we should devise a means of shifting (escaping) the unconscious state to address this limitation. This paper solves the limitation by shifting the unconscious state by connecting the conscious state to the unconscious state via the linear layer $f_\theta$ and gating function $f_\gamma$ as
\begin{equation}
    \label{eq:unconscious2conscious}
    h'_{l-1} = \gamma_l \odot f_\theta (c_l) + (1-\gamma_l) \odot h_{l-1},
\end{equation}
where the connection gate $\gamma_l$ is given by
\begin{equation}
    \label{eq:connectiongate}
    \gamma_l = f_\gamma\left(c_l, h_{l-1}\right) = \sigma\left(W_{\gamma} c_l+U_{\gamma} h_{l-1}\right).
\end{equation}
The unconscious state is instantaneously shifted when the conscious state is updated; thus, the conscious has an immediate impact on the unconscious.
Therefore, equation~(\ref{eq:odesolver}) should be modified as 
\begin{equation}
    \label{eq:odesolver_r}
    h_{l} =\mathrm{ODESolver}\left(\underline{h'_{l-1}}, {a}(u,t), t_{l-1}, t_{l}, \{m_i, e_i | i \in \mathcal{I} \}\right).
\end{equation}
Notably, we shift not only the position but also its velocity. This process can be regarded as an asteroid crash on a planet that changes a planet's velocity and position.

\subsubsection{Generating the decision-making state}
The proposed model generates a decision-making state $d_l$ at time $t_{l}$ by balancing the conscious state $c_l$ and unconscious state $u_l$ via the linear layer $f_\phi$ and gating function $f_\gamma$ as
\begin{equation}
    \label{eq:decision-making-state}
    d_{l} = \delta_l \odot f_\phi (c_l) + (1-\delta_l) \odot u_l,
\end{equation}
where the decision-making gate $\delta_l$ is given by
\begin{equation}
    \label{eq:delta}
    \delta_l = f_\delta (c_l, u_l)= \sigma\left(W_{\delta} c_l+U_{\delta}u_l\right).
\end{equation}
This process mimics Freud's reality principle: $\delta_l$ determines the importance of the conscious and unconscious states. 
When $\delta_l$ is large, the conscious state is more influential than the unconscious state;
in other words, the user is relatively rational. 
On the contrary, when $\delta_l$ is small, the unconscious state is essential for estimating consumption.

\subsection{Recommendation layer}
The proposed model recommends the nearest items to the decision-making state at time $t_{l}$ as
\begin{equation}
    \label{eq:recommend_item}
    s_{l} = \argmin_{i\in\mathcal{I}}{\norm{d_{l}-e_i}}.
\end{equation}
Sharing item embedding vectors (\ie, $e_l$) in the item-embedding layer and nearest item finder can reduce overfitting.
To train our model by backpropagation, we modify equation~(\ref{eq:recommend_item}) to be differentiable, as follows:
\begin{equation}
    \label{eq:item_prob}
    p_{li} = \frac{e^{-\norm{d_{l}-e_i}^2}}{\sum_{j} e^{-\norm{d_{l}-e_j}^2 }},
\end{equation}
where $p_{li}$ is the probability of item $i$ being recommended at time $t_{l}$. 
This equation is a form of the softmax function, which receives the negative squared Euclidean distance between the decision-making state and item embedding.

\subsection{Training of the proposed model}
We train the proposed model in an end-to-end manner. To learn the model parameters described in equations~(\ref{eq:item_emb}) to~(\ref{eq:odesolver}), we use a cross-entropy loss function as an objective function to be minimised:
\begin{equation}
    \label{eq:loss_function}
    \mathcal{L}=-\sum_{b=1}^{\mathcal{B}}\sum_{l,i} y_{li}^{b}\log p_{li},
\end{equation}
where $\mathcal{B}$ is the mini-batch size and $y_{li}^b$ is a binary indicator of whether item $i$ was interacted with at the $l$-th order in the $b$-th behaviour sequence.
Let $\mathcal{S}^b$ be the $b$-th behaviour sequence and $s_l^b$ and $t_l^b$ be the interacted item index and interaction time in the sequence, respectively. Then, $y_{li}^b$ has a value of one only when $i=s_l^b$ and is zero otherwise.
We obtain $p_{li}$ by passing $s_l$ through the embedding layer, \ours, and recommendation layer in consecutive order and train the whole model by backpropagating the loss.
This paper uses theAdam optimiser~\cite{kingma2015adam} with gradual warm-up learning rate scheduler~\cite{goyal2017accurate} on the mini-batch (see Supplementary Section~\ref{sec:sm:minibatching} for details on mini-batching strategy).

\section{Results}
\label{sec:exp}

\subsection{Recommendation performance}

Fig.~\ref{fig:performances} shows the recommendation performance of \ours~and four baselines (see Supplementary Section~\ref{sec:sm:baselines} for details on the baselines) on six real-world benchmark datasets (see Supplementary Section~\ref{sec:sm:dataset} for details on datasets).
From an economic perspective~\cite{nelson1970information,franke2004information}, we can classify the datasets into three product groups -- search product group, experience product group, and mixed product group.
The search product group is a set of goods whose characteristics are clearly evaluated before purchase, and the experience product group is a collection of goods whose characteristics are difficult to assess in advance but can be ascertained after consumption. Some goods in the mixed product group have characteristics of the search product group, and the remaining goods belong to the experience product group. 
The `clothing, shoes, \& jewelry' and `patio, lawn, \& garden' datasets belong to the search product group, and the `beauty' dataset belongs to the experience product group. 
The remaining datasets belong to the mixed product group. 
Our results demonstrate that \ours~is effective on the search product group, whereas the proposed model is less effective on the experience and mixed product groups. 
These results may be derived by modelling the unconscious state, and we further investigate the inherent mechanism of the proposed model.

\begin{figure}[t!]
    \centering
    \begin{subfigure}{0.7\textwidth}
        \centering
        \includegraphics[width=\textwidth]{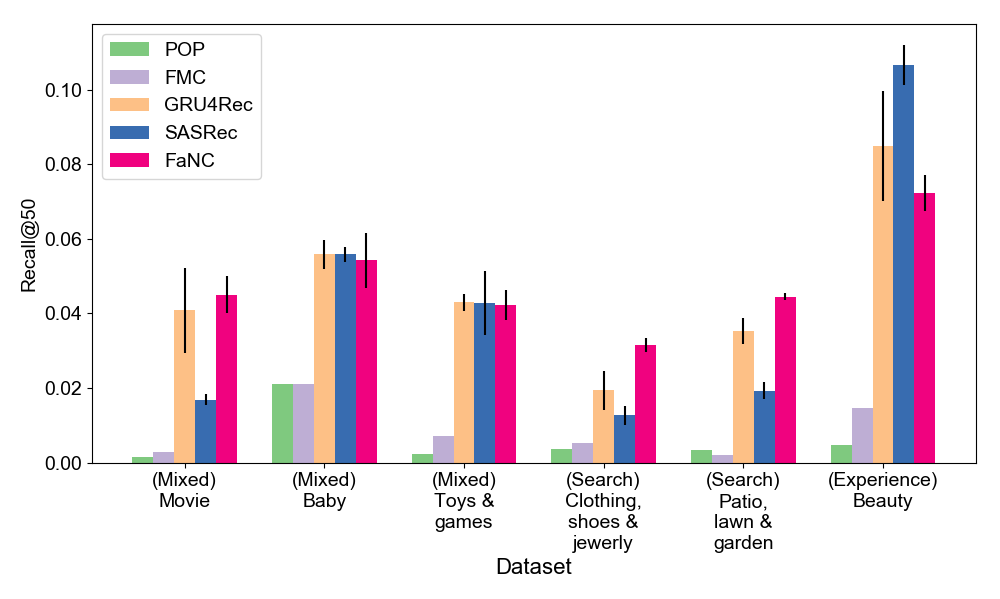}
    \end{subfigure} \par
    \begin{subfigure}{0.7\textwidth}
        \centering
        \includegraphics[width=\textwidth]{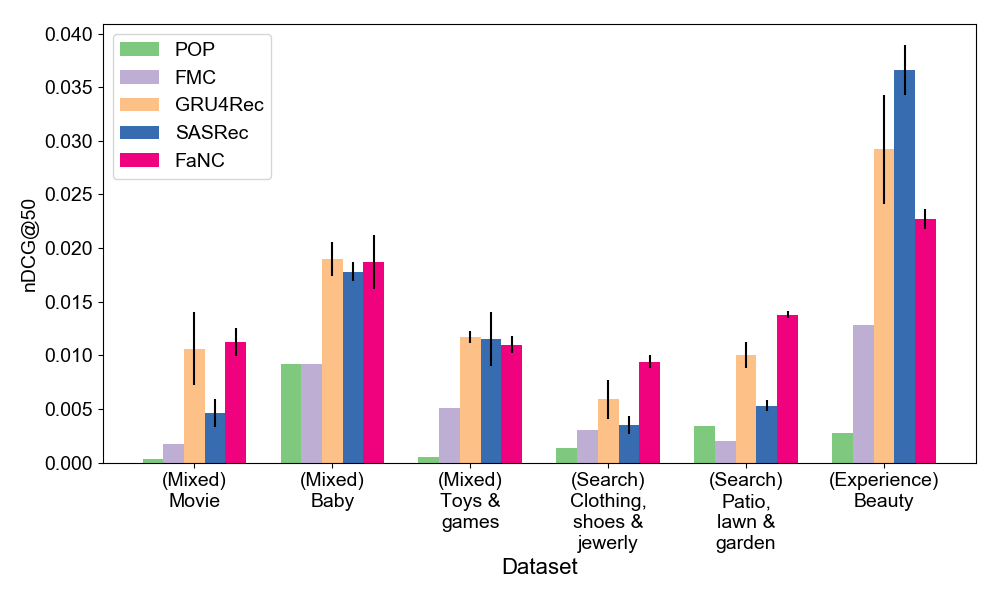}
    \end{subfigure}
    \caption[Performance of \ours~on real-world benchmark datasets.]{Performance of \ours~on real-world benchmark datasets. We evaluated \ours~on six datasets of three different groups -- search product group, experience product group, and mixed product group -- in terms of recall (top) and nDCG (bottom). 
    All values shown in the figures are the averages of five replicates. 
    }
    \label{fig:performances}
\end{figure}

Furthermore, we observed that the models' performance varies depending on the product group.
For the experience product group, SASRec recorded the highest performance, and ours showed the lowest performance. 
By contrast, the opposite result is observed for the search product group. 
One plausible explanation is that this pattern occurred due to the differences in statistics between product groups~(Supplementary Table~\ref{tab:dataset_sum}). As the datasets belonging to the search product group generally contain a large number of items, SASRec may have difficulty learning the representations of many items. However, \ours~achieved better results since it reflects human's inductive biases.

\subsection{Effects of the pleasure and reality principles}
\begin{figure}[t!]
    \centering
    \begin{subfigure}{0.3\textwidth}
        \centering
        \includegraphics[width=\textwidth]{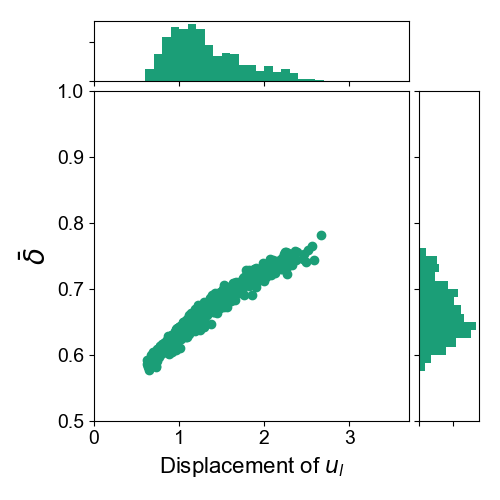}
        \caption{}
        \label{fig:scatter:movie}
    \end{subfigure} 
    \hfill
    \begin{subfigure}{0.3\textwidth}
        \centering
        \includegraphics[width=\textwidth]{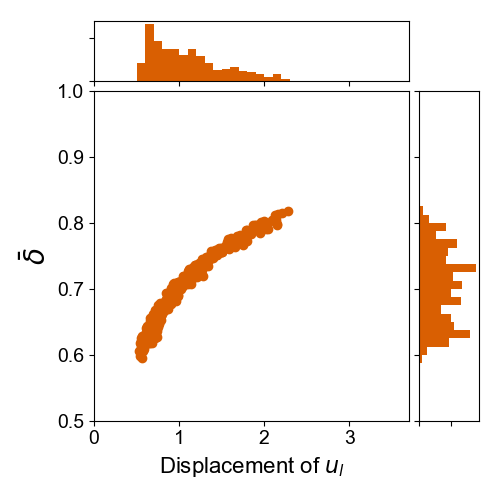}
        \caption{}
        \label{fig:scatter:baby}
    \end{subfigure} 
    \hfill
    \begin{subfigure}{0.3\textwidth}
        \centering
        \includegraphics[width=\textwidth]{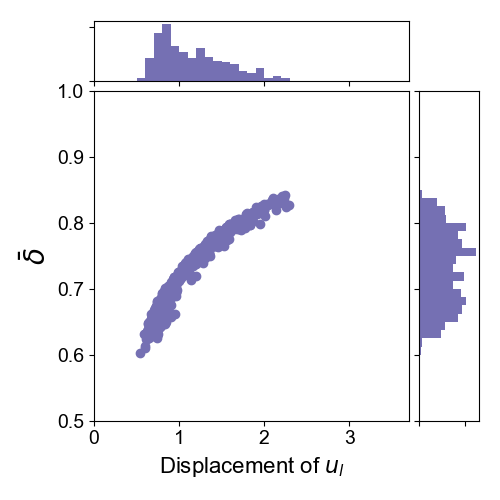}
        \caption{}
        \label{fig:scatter:toy}
    \end{subfigure} 
    \par
    \begin{subfigure}{0.3\textwidth}
        \centering
        \includegraphics[width=\textwidth]{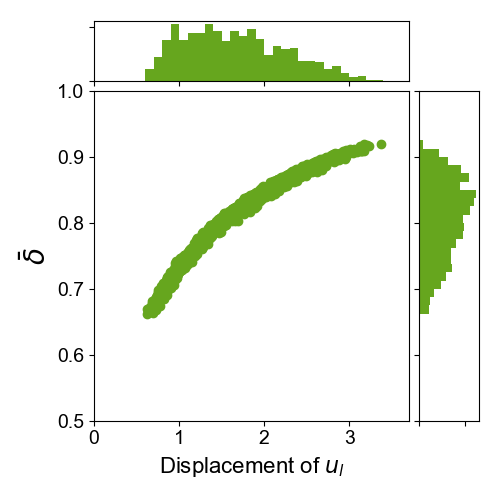}
        \caption{}
        \label{fig:scatter:clothing}
    \end{subfigure} 
    \hfill
    \begin{subfigure}{0.3\textwidth}
        \centering
        \includegraphics[width=\textwidth]{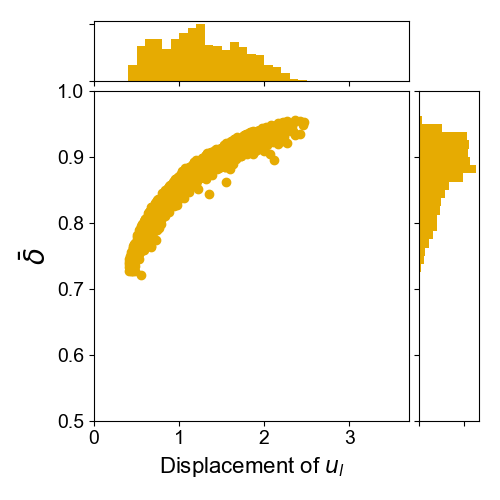}
        \caption{}
        \label{fig:scatter:patio}
    \end{subfigure} 
    \hfill
    \begin{subfigure}{0.3\textwidth}
        \centering
        \includegraphics[width=\textwidth]{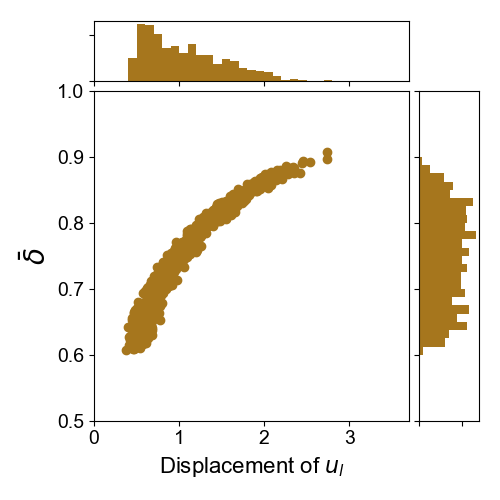}
        \caption{}
        \label{fig:scatter:beauty}
    \end{subfigure} 
    \caption[Scatter plot with marginal histograms of user behaviour via pleasure principle and reality principle.]{Scatter plot with marginal histograms of user behaviour via pleasure principle and reality principle on the movie~(\subref{fig:scatter:movie}), baby~(\subref{fig:scatter:baby}), toys \& games~(\subref{fig:scatter:toy}), clothing, shoes \& jewerly~(\subref{fig:scatter:clothing}), patio, lawn \& garden~(\subref{fig:scatter:patio}), and beauty~(\subref{fig:scatter:beauty}) datasets. The x- and y-axes represent the displacement of the unconscious state $u_l$ during the free-floating process (\ie, pleasure principle) and the average importance of the conscious state $\bar{\delta}$ when making a decision (\ie, reality principle), respectively. The importance ranges from zero to one: the conscious state is more influential than the unconscious state when the value is large.}
    \label{fig:inner_mechanism}
\end{figure}

Fig.~\ref{fig:inner_mechanism} shows how the pleasure and reality principles work in the trained proposed model. 
From the figure, we observed two characteristics of \ours.
First, the reality principle tends to suppress the unconscious state as the state floats according to the pleasure principle.
For all datasets, the proposed model tends to place more importance on the conscious state as the unconscious state is free-floating, even though users' unconscious states did not flow in the same direction.
Thus, users may be rational over the long term. In other words, the longer a user spends contemplating before buying a product, the more rational the user becomes.
Second, this paper also found that \ours~ behaves similarly to actual consumers.
The marginal distributions of importance (\ie, vertical distribution) of search product groups are upper skewed, and those of the other groups are lower skewed or uniform. 
Thus, our model increased the importance of the conscious state when providing recommendations on search product compared to that on other products. 
These results are in accordance with consumer behaviour, which is more dependent on the conscious than unconscious when buying search products (\ie, simple situation) compared to experience products (\ie, complicated situation)~\cite{dijksterhuis2006making,gao2012understanding}. 
The latter characteristic may also be responsible for the different results for the search product group and other groups.
Notably, we did not intend for our model to have these characteristics; we merely included the free-floating unconscious state in the model.

\subsection{Example what-if analysis of consumption time}
\begin{figure}[t!]
    \centering
    \begin{subfigure}{0.8\textwidth}
        \centering
        \includegraphics[width=\textwidth]{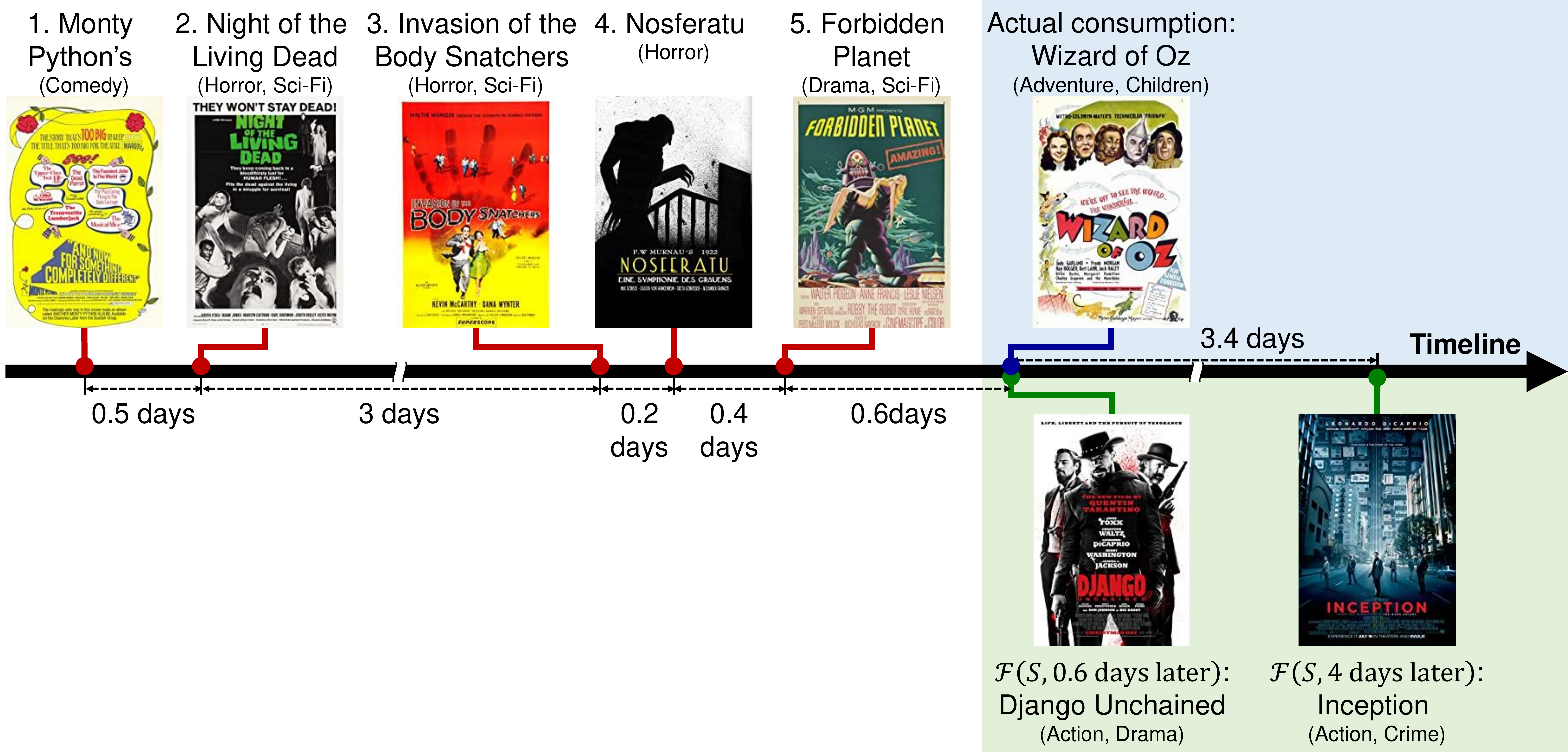}
        \caption{}
        \label{fig:whatif:1}
    \end{subfigure} \par
    \begin{subfigure}{0.8\textwidth}
        \centering
        \includegraphics[width=\textwidth]{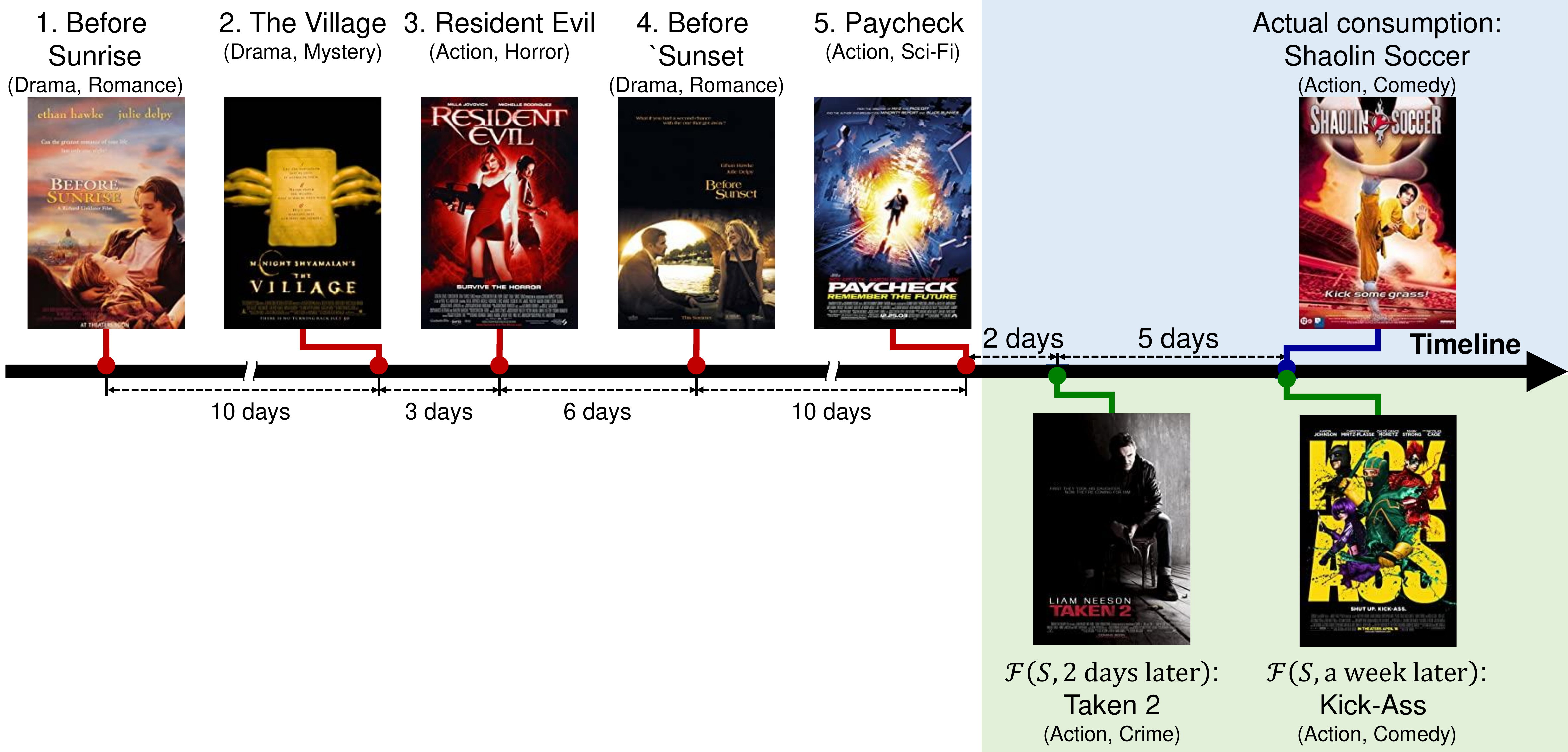}
        \caption{}
        \label{fig:whatif:2}
    \end{subfigure}
    \caption[Illustration of a what-if analysis of consumption time for two sample users on the movie dataset.]{Illustration of a what-if analysis of consumption time for two sample users on the movie dataset. The top-left five movies indicate the recently consumed items, and the top-right movie is the actually consumed item. The bottom-right two movies are the recommended movies at the corresponding time. (\subref{fig:whatif:1}) is the what-if analysis if the next item was consumed later than it actually was, and (\subref{fig:whatif:2}) is the what-if analysis if the next item was consumed earlier than it actually was.
    This figure shows the genre of each movie below the title.
    }
    \label{fig:whatifanalysis}
\end{figure}

Since this paper modified the definition of the sequential recommendation model as in the equation~(\ref{eq:problem}), \ours~can provide a what-if analysis of consumption time. We illustrate two examples of a what-if analysis of consumption time on the movie dataset in Fig.~\ref{fig:whatifanalysis}. 
The first user in Fig.~\ref{fig:whatif:1} had seen gloomy movies before actually consuming the item \emph{wizard of oz}. This example shows that \ours~fails to successfully recommend an item when users' preferences change dramatically: our model recommended dark types of movies that are similar but slightly different. 
The second user in Fig.~\ref{fig:whatif:2} alternated between romantic and dark movies, and \ours~recommended a crime movie and an action-comedy movie.
The results imply that time might be an important factor because the user's unconscious evolves.
Therefore, the effectiveness of recommendation can be maximised if the service provider considers the time, such as the release date~\cite{chiou2008timing,lee2014pricing} and advertising time~\cite{feichtinger1994dynamic}, to maximise profit.

\subsection{Discussion}
\label{sec:dis}
Even though the proposed model reflects the users' unconscious state, \ours~follows Freud's unconscious in a complete sense.
Freud believed that the unconscious determines individual behaviour. 
Our modelling lets users' unconscious states lie in the same latent space as the item embeddings, which means that \ours~ models users' common behaviour patterns.
By contrast, Jung, a psychoanalyst as famous as Freud, claimed that the unconscious can be divided into a collective unconscious and personal unconscious~\cite{jung1916structure}; thus, \ours~can be regarded as reflecting the collective unconscious, which does not develop individually but is inherited because we are human.
To model a single user's individual behaviour pattern from the Freudian perspective, we must consider the personal unconscious, which develops individually.
We expect meta-learning methods~\cite{vinyals2016matching,snell2017prototypical,finn2017model} to be candidates for modelling personal unconscious because they can produce a personalised item embedding space or decision-making process based on consumed items~\cite{pan2019warm,lee2019melu}.
Considering the personal unconscious via meta-learning may improve the performance not only on the search product group but also on the experience product group.

Another interesting direction to improve our model exists. We assumed that only the user's unconscious state is floating and that the item embeddings are fixed in the latent space. These assumptions were made because the $N$-body problem, i.e., predicting the individual motions of many particles interacting with each other gravitationally, is intractable in physics~\cite{hemsendorf2002instability}, and placing items in the latent space made our model tractable.
We believe that it is possible to make the $N$-body problem in the recommender system tractable by using advanced machine learning methods in the future.

\section{Conclusion}
\label{sec:con}
This paper proposed a recurrent cell, namely, \ours, for the sequential recommendation from the Freudian and Newtonian perspectives.
From the Freudian perspective, we divided a user's state into conscious and unconscious states and modelled the user's conscious and unconscious decision-making process by means of the pleasure and reality principles.
To model the pleasure principle, \ie, the user's free-floating unconscious, this paper followed the Newtonian perspective. We modelled the reality principle via a gating function that balances the conscious and unconscious states.
The proposed model outperformed baseline methods on datasets belonging to the search product group and produced comparable results on the other datasets. 
\ours~opens a new direction of sequential recommendation at the convergence of psychoanalysis and recommender systems that enables us to understand users' decision-making processes.


\section*{Data availability}
The movie (MovieLens20M) dataset is publicly available at \url{https://grouplens.org/datasets/movielens/} and other datasets are is publicly available at \url{https://jmcauley.ucsd.edu/data/amazon/}.

\newpage
\beginsupplement

\section*{Supplementary Materials}
\section{Mini-batching strategy for \ours}
\label{sec:sm:minibatching}
\begin{figure*}[h!]
    \centering
    \includegraphics[width=0.7\textwidth]{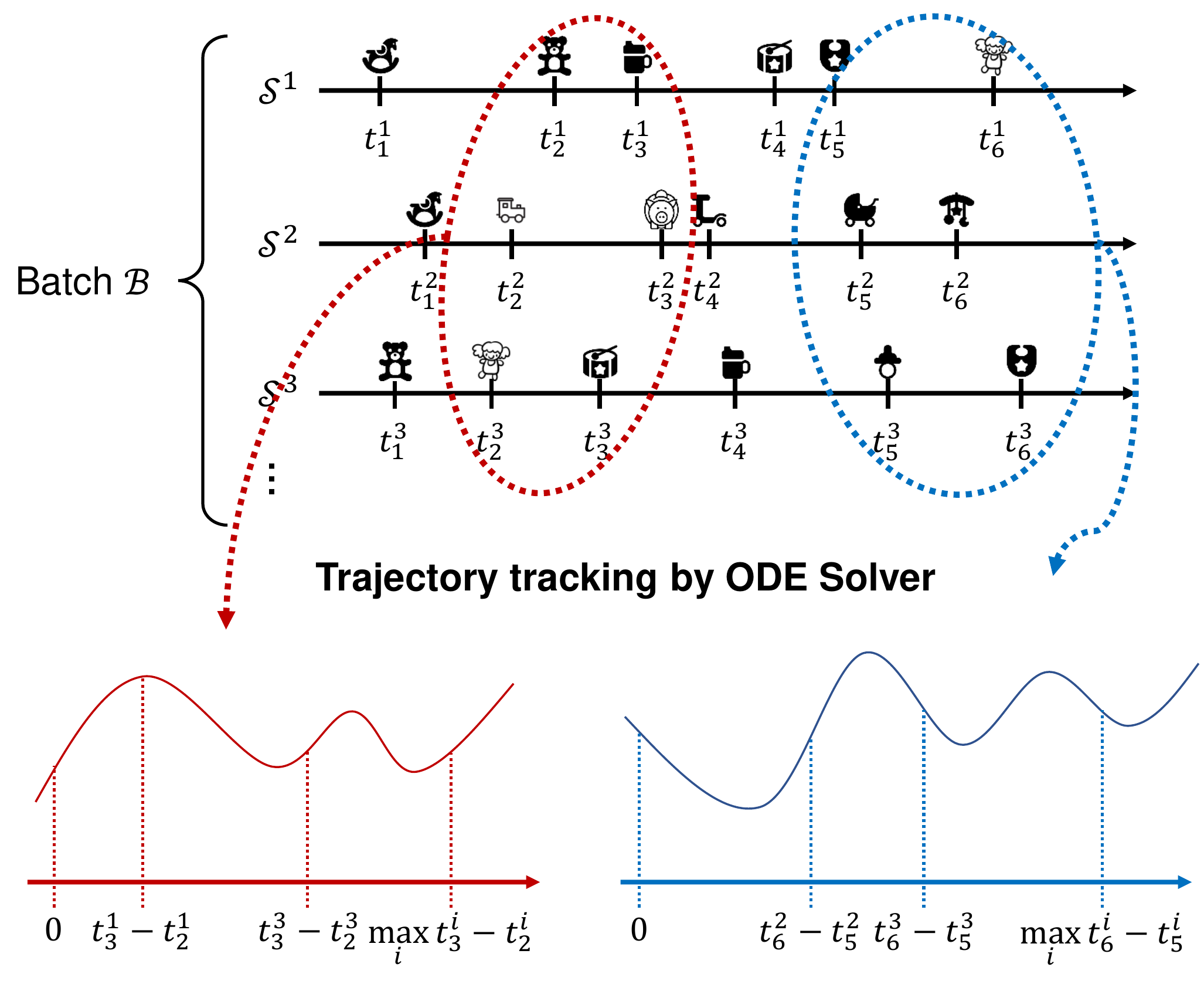}
    \caption[Illustration of the mini-batch trick]{Illustration of the mini-batch trick. To enable the mini-batch strategy for \ours~ during training, we trace the floating paths to obtain the next states from users with different time intervals. After tracing, the model assigns the unconscious state of the corresponding time interval in the path to the user's next unconscious state.}
    \label{fig:batch_trick}
\end{figure*}

To allow mini-batching under ODE, we utilise the trajectory tracking of the ODE solver to obtain the next states from multiple users who have different time intervals, as shown in Figure~\ref{fig:batch_trick}. 
The neural ODE has a drawback in that it is difficult to train the model in a mini-batch-wise manner because the ODE solver typically flows instances (\ie, unconscious states) in the mini-batch over the same amount of time. 
Since users' item-consumption patterns are diverse, we should consider a mini-batch that has varied time intervals (\textit{e.g.}, $t_3^1-t_2^1\neq t_3^2-t_2^2$ in the figure).
Therefore, we group items with the same sequence order within the mini-batch and calculate the mini-batch states sequentially according to the order.
Because equation~(13) holds, only the time intervals between the $l-1$-th and $l$-th order are required to calculate the $l$-th mini-batch states.
To handle different time intervals in a mini-batch, we let all unconscious states float freely during the maximum time interval (which we call time padding) and track the floating paths.
Then, we assign the state of the corresponding time interval in the path to each instance's next state.
This approach is one of the simplest ways to apply mini-batching, and the time padding is similar to the zero padding for the mini-batch in NLP models~\cite{cho2014learning_s,hu2014convolutional}.

\section{Experimental details}
\label{sec:sm:expdetail}

\subsection{Comparison methods}
\label{sec:sm:baselines}
We compared the proposed model with four baseline models. 
Two baselines were classic models, and the others were NLP-based models for sequential recommendation. Specifically, we employed the following models:

\begin{itemize}
    \item Classic recommendation models:
    \begin{itemize}
        \item POP: This model, the simplest baseline, ranks items in descending order in terms of the popularity calculated based on the number of user-item interactions in the training set.
        \item Factorised Markov chain (FMC): Following the first-order Markov assumption, this model ranks items according to transition probability given the item in the last action, which is estimated in the training set.
    \end{itemize}
    \item NLP-based sequential recommendation models:
    \begin{itemize}
        \item GRU4Rec~\cite{hidasi2015session_s}: This model projects items in the embedding space and uses GRU to capture user behaviour dynamics based on the embedded items and to recommend items. 
        \item SASRec~\cite{kang2018self_s}: This model uses left-to-right unidirectional self-attention layers (\ie, Transformer~\cite{vaswani2017attention_s}) to capture user behaviour dynamics based on the embedded items and recommends items in a similar way to GRU4Rec. 
    \end{itemize}
\end{itemize}
We did not consider collaborative-filtering-based models because they are unsuitable for predicting items under the typical sequential recommendation setting.

\subsection{Dataset}
\label{sec:sm:dataset}
We considered six public benchmark datasets from real-world applications: MovieLens 20M (movie for short) ~\cite{harper2015movielens_s}, Amazon baby, Amazon beauty, Amazon clothing, shoes, \& jewelry, Amazon patio, lawn \& garden, and Amazon toys \& games~\cite{mcauley2015image_s}.
The Movielens 20M dataset is one of the most widely used datasets to evaluate recommendation models. 
The Amazon datasets are the corpora of product reviews crawled from the online shopping platform \textit{Amazon.com}. 

According to previous studies on information economics~\cite{franke2004information_s,bae2011product}, the majority of products in the `clothing, shoes, \& jewelry' and `patio, lawn \& garden' datasets are search products.
Some of the products in `baby' and `toys \& games' datasets are search products and the rest are experience products. 
Moreover, most of the products in the `beauty' and `movie' datasets are experience products.
However, `movie' also has characteristics of experience products. 
As over-the-top services such as Netflix emerge, users are selecting films based on the their thumbnails~\cite{amat2018artwork}. 
Thus, the thumbnail can provide partial information about the characteristics of a movie to users and can be used by users to ascertain movies.
Thus, the `movie' dataset may show a similar tendency to the consumption of the mixed search and experience products such as `baby' and `toys \& games'.
These characteristics yield different results because consumers have different consumption patterns~\cite{luan2016search,yang2010experiential}.

This paper converted all datasets into the sequential recommendation setting, as in the previous work~\cite{kang2018self_s,rendle2009bpr}.
We selected only behaviour sequences that have different item interaction timestamps. 
Although previous research~\cite{kang2018self_s,rendle2009bpr} used timestamps to determine the sequence order of item interaction, they did not provide information on how they decided the order when the items have the same timestamp. 
Since using behaviour sequences with the same timestamp derive the recommendation model to train alphabetic order or numerical order of the items' identification rather than the user's sequential behaviour pattern, we removed these sequences.
Moreover, we set the time unit of the ODE to a week for the `movie' dataset and to three months for the other datasets. To resolve the memory issue of ODE, this paper limited the maximum time interval to 1.5 units of time, \ie, the time padding was set to 1.5 time units.
Then, we divided the training, validation, and test datasets based on the sequence's identification. We randomly selected eighty percent of the sequences for training, ten percent for validation, and the rest for testing. 
Notably, we used the all items except the actually consumed item as negative samples.
Therefore, we did not conduct a negative sampling of items for the prediction, as in previous papers~\cite{kang2018self_s,he2017neural}, because that approach may cause the performance to be overestimated compared to the actual recommendation environment, where sampling is difficult to apply. 
Table~\ref{tab:dataset_sum} shows the statistics of the datasets after selecting the sequences.

\begin{table}[t!]
    \centering
    \begin{tabular}{l|c|c|c}
        \hline
         Dataset & \makecell{Number \\ of items}& \makecell{Number of \\training \\sequences} & \makecell{Average \\ time interval} \\
         \hline
         Movie & 27,278 & 502 & 1.27 weeks \\
         Baby & 7,050 & 928 & 4.86 months \\
         Toys \& Games & 11,924 & 1,487 & 5.21 months \\
         Clothing, Shoes, \& Jewelry & 23,033 & 3,103 & 4.38 months \\
         Patio, Lawn \& Garden & 101,902 & 5,672 & 7.39 months \\
         Beauty & 12,101 & 1,883 & 4.57 months \\
         \hline
    \end{tabular}
    \caption{Statistics of datasets.}
    \label{tab:dataset_sum}
\end{table}

\subsection{Evaluation measures}
The performance indicators used in this study were the recall at $k$ (Recall@$k$) and normalised discounted cumulative gain at $k$ (nDCG@$k$).
The recall at $k$ measures whether an item is recommended correctly, and it is calculated as
\begin{equation}
    \label{eq:recall}
    \mathrm{Recall}@k = \begin{cases}
    1,  & \text{if actually consumed item is in the top-$k$ list},\\
    0,  & \text{otherwise}.
    \end{cases} \\
\end{equation}
Additionally, we used the nDCG at $k$ to measure the quality of ranking. This metric has a large value when ranking a recommended item at the top rank and a low value otherwise. The definition of nDCG at $k$ is as follows:
\begin{align}
    \label{eq:ndcg}
    \mathrm{nDCG}@k &= \frac{DCG_k}{IDCG_k}, \\
    \mathrm{DCG}_k &= \sum_{l=1}^L \frac{2^{R_{l}}-1}{\log_2(1+l)}
\end{align}
where $R_{l}$ and $\mathrm{IDCG}_k$ are the relevance of the $l$-th ranked item and the best possible (\textit{i.e.} ideal) $\mathrm{DCG}_k$, respectively. 
This paper provides the average value of the evaluation measure of $L$ items in all sequences after five replicates.

\subsection{Implementation details}
For \ours, the embedding sizes of the item and unconscious state were both 16, and the dimension of the conscious state was 8. 
We used the Runge-Kutta method-based~\cite{butcher1987numerical} ODE solver.
The mini-batch size was set to 4 because the ODE solver requires substantial memory allocation. 
The sequence length $L$ was set to 10 for the movie dataset and 5 for the five other datasets. 
The learning rate was optimized by grid search for each dataset, and it was set to 0.0001 for all datasets. 
We implemented the proposed model with the torchdyn library~\cite{poli2020torchdyn} written in PyTorch~\cite{paszke2019pytorch}.
For GRU4Rec and SASRec, we set the embedding size of items to 16 and that of the hidden state to 8 to make the capacity of the models similar.
For all models, we set the maximum number of epochs to 100 and applied an early stopping strategy~\cite{yao2007early} when no improvement in validation loss was observed for ten consecutive epochs.


\end{document}